**Noncollinear enhancement cavity for record-high out-coupling efficiency of an extreme-UV frequency comb**

Chuankun Zhang[1], Stephen B. Schoun[1], Christoph M. Heyl[1,*], Gil Porat[1,†], Mette B. Gaarde[2], Jun Ye[1]

[1]JILA, National Institute of Standards and Technology and Department of Physics, University of Colorado, Boulder, Colorado 80309, USA

[2]Department of Physics and Astronomy, Louisiana State University, Baton Rouge, Louisiana 70803 USA

Abstract:

We demonstrate a femtosecond enhancement cavity with a crossed-beam geometry for efficient generation and extraction of extreme-ultraviolet (XUV) frequency combs at a 154 MHz repetition rate. We achieve a record-high out-coupled power of 600 µW, directly usable for spectroscopy, at a wavelength of 97 nm. This corresponds to a >60% out-coupling efficiency. The XUV power scaling and generation efficiency are similar to that achieved with a single Gaussian-mode fundamental beam inside a collinear enhancement cavity. The noncollinear geometry also opens the door for the generation of isolated attosecond pulses at >100 MHz repetition rate.

Frequency combs spectrally covering the extreme-ultraviolet (XUV) region have demonstrated, and continue to promise, prominent scientific advances in precision spectroscopy and attosecond physics. Direct XUV frequency-comb spectroscopy [1] of few-electron systems can provide stringent tests on quantum electrodynamics [2]. By opening up the vast spectral range of XUV for high-resolution spectroscopy, the feasibility of an optical clock based on a nuclear transition in [229]Th [3,4] driven with an XUV frequency comb is growing rapidly. In the time domain, the high repetition rate and ultrafast characteristics of XUV frequency combs allow time-resolved studies of dynamics in molecular and solid-state systems on femtosecond and even attosecond timescales with superior data acquisition speed and high signal-to-noise ratio [5–7].

A dispersion-managed passive optical cavity, known as a femtosecond enhancement cavity (fsEC) [8,9], is the key enabling technology for XUV frequency-comb generation [10,11]. A femtosecond pulse train, usually in the infrared (IR), can be coherently enhanced inside an fsEC, resulting in intracavity femtosecond pulses with >100 µJ pulse energy at multi-ten-MHz repetition rate [12]. The high pulse energy enables efficient frequency up-conversion via high-harmonic generation (HHG). At the same time, the coherence property of the fundamental drive laser is fully transferred to the XUV at the frequency comb's original repetition rate [1,13].

It has been challenging, however, to extract the intracavity-generated XUV power from an fsEC efficiently for scientific applications [14]. Various methods have been experimented with varying degrees of success in existing designs. Intracavity Brewster plates offer out-coupling efficiencies (OCEs) ranging from 5%-15%, depending on the material used and the harmonic order of interest [15]. Coated plates can offer up to 75% OCE at 149 nm [16], but suffer from relatively rapid degradation under XUV irradiation. An intracavity plate also introduces dispersive, nonlinear, and thermal effects, limiting power scalability. A cavity mirror with a nanograting etched on its top layer has been used to diffract the XUV light out with ~10% efficiency, while remaining as a high reflector for the fundamental beam [17]. Hydrocarbon buildup in the nanograting structure causes degradation of its efficiency with high XUV flux, which can be partly mitigated if immersed in an ozone-rich environment [12]. A few variations on the reflection-based out-coupling methods have been proposed, but not yet experimentally tested [18,19]. On-axis pierced mirrors offer direct access of the XUV light [14]. Experimental results and accompanying simulations show that 5% [20,21] to potentially 20% OCE [22,24] can be achieved using a Gaussian cavity mode. In order to improve the OCE with pierced mirrors, specially tailored higher-order spatial modes have been explored for HHG [14,23–25], however, to date the out-coupled power is lower than that achieved from regular Gaussian modes [12].

Cavity-enhanced noncollinear HHG was proposed in the early stages of fsEC development for efficient extraction of the generated harmonics [14,26,27]. The harmonics generated by two crossed beams are naturally separated from the fundamental at the bisection angle, and can thus be coupled out from the cavity geometrically, while the fundamental is recycled to maintain a high cavity buildup. Such a noncollinear geometry also offers unique opportunities for studying and controlling the HHG process in single-pass experiments. Since the early proposal and demonstration [28,29], single-pass noncollinear HHG has been implemented for generating circularly polarized XUV beams [30,31], gating isolated attosecond pulses [32,33], studying phase-matching processes [34,35], and for fundamental studies of extreme-nonlinear optics [36,37].

In this Letter, we report the development of a unique cavity geometry that allows two laser pulses to be resonantly enhanced simultaneously [Fig. 1(a)]. The two pulses overlap both spatially and temporally exactly at the cavity focus. We employ a small noncollinear angle in order to optimize the harmonic beam profile while avoiding a large phase mismatch imposed by the noncollinear geometry. Harmonic orders of 9-19 are measured. The out-coupled 11$^{th}$ harmonic reaches a record-high average power of 600 µW, which is 5 times higher than previously reported values [12]. This work establishes a powerful tool for delivering XUV frequency comb to spectroscopy targets, and represents an important step towards noncollinear gating in optical cavities for attosecond physics [32,33,38].

As schematically shown in Fig. 1(a), our experiment employs a 120 fs, 154 MHz repetition rate Yb:fiber frequency comb [39] with up to 80 W average power, spectrally centered at 1070 nm, to coherently seed an fsEC. The intracavity light field is linearly polarized perpendicular to the cavity plane. The cavity free spectral range is set at 77 MHz, resulting in two pulses circulating simultaneously inside the fsEC. A pinhole (not shown) is positioned at the focus to ensure the spatial overlap of the two pulses when the cavity is being aligned, and is removed during HHG operation. The temporal delay between the two laser pulses at the cavity focus is controlled with a piezoelectric actuator mounted on one of the out-coupling mirrors, and is intrinsically stable thanks to the large fraction of shared parallel optical paths in the cavity. No noticeable drift of the relative phase is observed during operation times of tens of minutes. Before the nonlinear medium is introduced, a single-beam power enhancement factor of ~170 is obtained inside the cavity. With a focal spot size $w_0 = 44$ µm ($1/e^2$ intensity radius), a peak intensity of $8 \times 10^{13}$ W/cm$^2$ is reached when the two pulses interfere constructively at the focus. A homemade glass nozzle wrapped with heater wires and with an orifice diameter of 50 µm [40], oriented perpendicular to the cavity plane, is used to inject the nonlinear medium (pure Xe or He:Xe mixture) to the cavity focus. Generated harmonics are coupled out through the gap between the two curved high-reflectivity mirrors. The 11th harmonic is directed to a NIST-calibrated detector. Transmitted IR light from one mirror is used for monitoring the intracavity power, mode profile, and pulse duration.

The two crossed beams form an intensity grating across the focal plane, see Figs. 1(b) and 1(c). Consequently, the beam profile of the harmonics is determined by the ratio $\eta = \beta/\gamma$ between the fundamental noncollinear half-angle $\beta$ and the Gaussian-beam divergence half-angle $\gamma$. For $\eta \gg 1$, XUV photons are generated at discrete angles dictated by photon energy-and-momentum conservation [36]. In the wave picture, interference between the harmonics generated by different fringes at the focus causes the angular separation of the harmonics in the far field [34]. As we reduce $\eta$ gradually, the far-field harmonics start to overlap and eventually merge together, as shown in the insets of Fig. 2. This occurs as significant harmonic power is generated only from the central fringe for sufficiently small $\eta$. For applications requiring undistorted unidirectional emission of harmonics, it is therefore important to keep $\eta$ small. On the other hand, clipping loss on the mirror edges increases dramatically as $\eta$ decreases to ~2. This effect reduces the cavity finesse and the power-buildup factor, thereby limiting the smallest useful $\eta$. For a given focal spot size $w_0$ (and thus $\gamma$), the angle $\beta$ and the gap size $d$ between the two mirrors determine both the power enhancement factor of the cavity and the OCE, as illustrated in Fig. 2.

The size of $\beta$ is important for phase-velocity matching between the harmonics and the fundamental. In addition to the usual neutral and plasma dispersion, one can show that a geometric wave vector mismatch arises from the noncollinear geometry, given in the small-angle approximation ($\beta \ll 1$) by

$$\Delta k_{nc}^q \approx \Delta k_c^q (1 + \frac{\pi \beta^2 z_R}{\lambda} - \frac{\beta^2}{2})$$

for harmonic order $q$, where $\Delta k_c^q \approx -\frac{q}{z_R}$ is the Gouy phase mismatch from a single Gaussian beam [34,35]. Here, $\lambda$ and $z_R$ are the wavelength and Rayleigh length of the fundamental beam, respectively. $\Delta k_{nc}^q$ can be compensated by a below-critical-ionization generation medium, in which dispersion from neutral atoms dominates over that from plasma. The intensity-dependent dipole phase of HHG can be neglected for the overall phase-matching consideration as the gas nozzle is placed very close to the focus and as the generation medium is much shorter than the Rayleigh length in our experiment [41]. However, it is advantageous to keep the geometric phase mismatch small in the first place. This is because, in fsECs, nonlinearities from the gas target disturb the resonant condition between the laser and cavity and cause transverse-mode coupling, resulting in a clamping effect on the intensity buildup of the fundamental beam [42,43]. A smaller $\Delta k_{nc}^q$ would require a lower phase-matching pressure. This allows us to operate the enhancement cavity in a regime with a lower gas density and thus a reduced intensity-clamping effect. To simultaneously obtain a useful cavity buildup, a uniform beam profile, a good OCE, and a small phase mismatch, optimal experimental conditions are achieved with $d$ = 0.5 mm and $\beta$ = 0.94° (see Fig. 2).

We perform numerical simulations to understand the harmonics generated in the crossed beams at the peak of the laser pulse. We calculate the HHG response in the plane of the laser focus using the intensity-dependent dipole amplitude and phase, predetermined from the solution of the time-dependent Schrödinger equation for a large intensity range [44–46]. We consider only the short-trajectory contribution, which is extracted from the dipole data by numerical filtering [47–52], and all other phase-matching effects are neglected in the simulation by taking into account HHG emission from the focal plane only. The harmonics generated are then propagated to the far field, including diffraction from the mirror edges, using Huygens's integral in the Fresnel approximation [53]. A peak intensity of $5 \times 10^{13}$ W/cm² is used in the simulation, close to the experimental laser intensity at the optimal generation condition. The harmonic beam shapes at the out-coupling-mirror surface plane (10.3 cm away from the focus) and in the far field (70 cm behind the out-coupling mirrors) are shown in Fig. 3. The relative carrier phase $\Delta\phi$ between the two pulses of the crossed beams changes the laser interference pattern at the focus [Fig. 1(c)], as well as the far-field harmonic profile. When $\Delta\phi = \pi$, the harmonics generated from different parts of the fundamental interfere destructively on the bisection axis. This causes the harmonic beam to split into a doublet in the far field. Experimentally observed harmonic spatial profiles are shown in the insets. Harmonics of order 9 to 19 are observed on a fluorescent plate (sodium salicylate), recorded in Fig. 4. The asymmetry in the experimentally recorded beam profile is caused by a slight misalignment between the bisection axis of the crossed beams and the center of the mirror gap. Theoretically estimated OCEs for these harmonic orders are shown in Fig. 4(c).

When studying the output XUV power in the 11$^{th}$ harmonic as a function of the intracavity fundamental power, we observe two counter-intuitive behaviors [Fig. 5(a)]. First, with pure Xe as the generation medium (green traces), the XUV output is higher when $\Delta\phi = \pi$ for the same fundamental drive power. Second, with a He:Xe mixture as the generation medium (purple traces), the XUV output is higher when $\Delta\phi = 0$ for the same fundamental power. But still, the highest XUV power available is obtained when $\Delta\phi = \pi$ due to its higher intracavity power. Further study shows that the seemingly surprising behaviors can be understood simply as a result of changing the focal volume shape. As shown in Fig. 1(c), $\Delta\phi$ changes the intensity grating at the focus and, therefore, the peak intensity. Since HHG is an extremely nonlinear process, most of the harmonic power is generated from the central peak for $\Delta\phi = 0$, or the two innermost peaks for $\Delta\phi = \pi$. The contributions from side peaks are negligible due to their weak intensities. We therefore refer to the volume of the central peaks as an *effective* generation volume. For our angle ratio $\eta = 2.13$, when we change $\Delta\phi$ from 0 to $\pi$, the fundamental power concentrated in the effective generation volume increases from 57% to 88%. In other words, a larger fraction of the fundamental power is contributing efficiently to the HHG process when $\Delta\phi = \pi$. We then determine an effective conversion efficiency as the ratio between the generated XUV power and the fundamental power in the *effective* generation volume. Remarkably, this effective conversion efficiency is approximately identical for $\Delta\phi = 0$ and $\Delta\phi = \pi$ through the entire range of peak intensities measured for each medium, as demonstrated in Fig. 5 (b).

Owing to the high repetition rate, the plasma generated from one laser pulse does not clear the focal volume before the next pulse comes in, resulting in a highly dispersive accumulated plasma in the generation volume that degrades phase matching. As shown in Ref. [12], the harmonic yield limited by the accumulated plasma is characterized by a dimensionless parameter $\xi_{beam} = \frac{\sigma_{FWHM}}{v_{gas}} \times f_{rep}$, which represents the number of laser pulses that one atom "sees" during its transit through the laser beam. Here $\sigma_{FWHM}$ is the full-width at half maximum of the focus and $v_{gas}$ is the average atom velocity. Following Ref. [12], we used a 9:1 He:Xe mixture heated to about 560 °C as the generation medium, corresponding to $\xi_{beam} \sim 5$ at our laser repetition rate of 154 MHz. We observed a significant gain in the harmonic yield, compared to using pure Xe, due to both reduced neutral-depletion and improved phase-matching conditions, see Fig. 5. Cavity bistability caused by a phase shift from the steady-state plasma is observed when we sweep the cavity over its resonance with the comb [42], shown in the inset of Fig. 5 (b). This indicates that a significant plasma density remains even when using the heated gas mixture. Further improvements in the harmonic conversion efficiency is anticipated with further reduction of the steady-state plasma until reaching $\xi_{beam} < 2$, where the cavity resonance will show a nearly Lorentzian lineshape.

With this successful demonstration of a dual-pulse noncollinear fsEC for efficient XUV frequency-comb generation and extraction, we now understand the phase-matching conditions and the HHG efficiency via crossed beams inside the cavity. The achieved record-high out-coupled XUV frequency-comb power will be directly applied to high-resolution XUV spectroscopy, including the search for the $^{229}$Th nuclear transition. Besides precision spectroscopy, ultrafast time-resolved studies with isolated attosecond pulses will also benefit from these results. With properly chosen pulse duration and delay, interferences between the two overlapping pulses will create an ultrafast wave-front rotation that streaks the generated attosecond pulses into different directions [32,33,38]. The noncollinear cavity is also compatible with advanced control of mirror dispersion [54] and nonlinear intracavity dynamics [55] to reduce the pulse duration.

The work is supported by AFOSR Grant No. FA9550-19-1-0148, NSF Grant No. PHY-1734006, and the National Institute of Standards and Technology. M. G. acknowledges support from NSF Grant No. PHY-171367. We thank L. v. d. Wense for technical assistance and S. Diddams for reading the manuscript.

[*]Present address: German Electron Synchrotron DESY, Notkestr. 85, 22607 Hamburg, Germany & Helmholtz-Institute Jena, Fröbelstieg 3, 07743 Jena, Germany

[†]Present address: Department of Electrical and Computer Engineering, University of Alberta, Edmonton, Alberta T6G 1H9, Canada

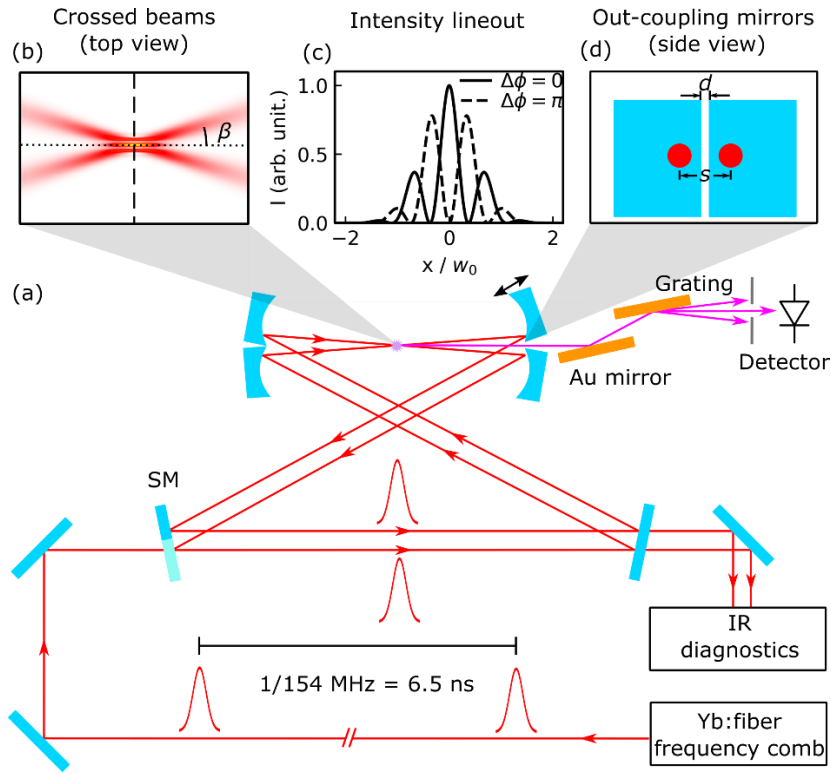

Fig 1. (a) Schematics of the noncollinear enhancement cavity. A high-power Yb:fiber frequency comb with 154 MHz repetition rate seeds a dual-pulse fsEC whose free spectral range is set to 77 MHz. The cavity is composed of 6 mirrors: four identical curved mirrors (radius of curvature: 20 cm), one flat mirror, and one segmented mirror (SM) which is home-made by bonding a high reflector and a 1.5% transmission input coupler side-by-side to a flat substrate. HHG is performed at the cavity focus, where the two circulating pulses overlap temporally and spatially. The temporal delay between the two pulses is controlled via a piezo-actuated mirror (indicated by black arrows). (b) Top view of the crossed beams (not to scale). Each beam forms an angle $\beta$ with the bisection axis (dotted line). (c) Intensity lineout at the dashed line in (b). The relative carrier phase $\Delta\phi$ between the two beams changes the interference pattern. $w_0$ is the beam waist. (d) Side view of the out-coupling mirrors. The two crossed beams, separated by a distance $s$ on the out-coupling-mirror surface, are aligned close to the edge of the mirrors with a gap size $d$.

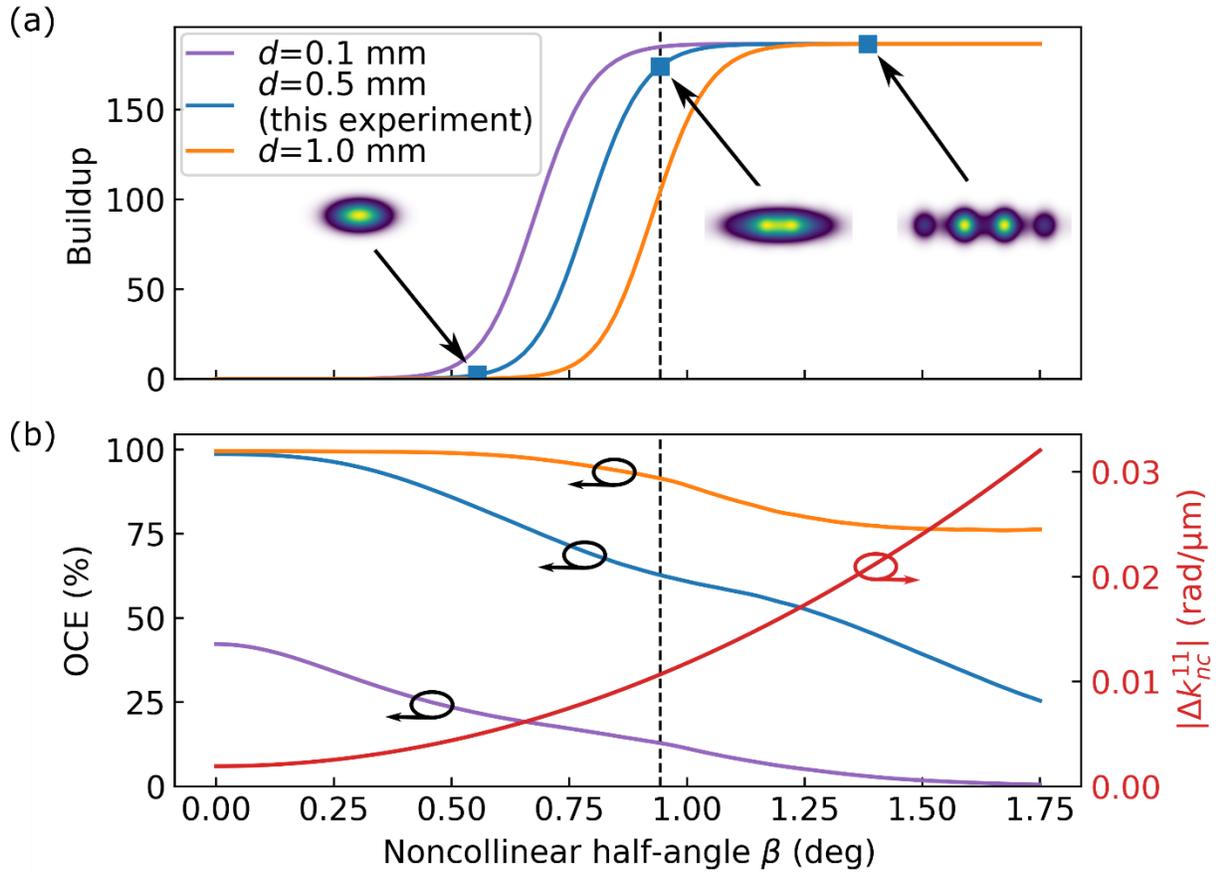

Fig 2. (a) Cavity buildup factor as a function of noncollinear half-angle $\beta$, shown for different mirror gaps $d$. Insets show the simulated 11$^{th}$ harmonic far-field distribution, immediately before the out-coupling mirrors. For large $\beta$, the harmonics split into separated spots. A 70% cavity mode-matching factor is assumed for the buildup calculation. (b) The out-coupling efficiency (OCE) for the 11$^{th}$ harmonic with different $d$, and the geometrical phase mismatch of the 11$^{th}$ harmonic $\Delta k_{nc}^{11}$, as a function of $\beta$. Our experimental conditions are $d$ = 0.5 mm (blue line) and $\beta$ = 0.94° (indicated by the dashed vertical line in both panels).

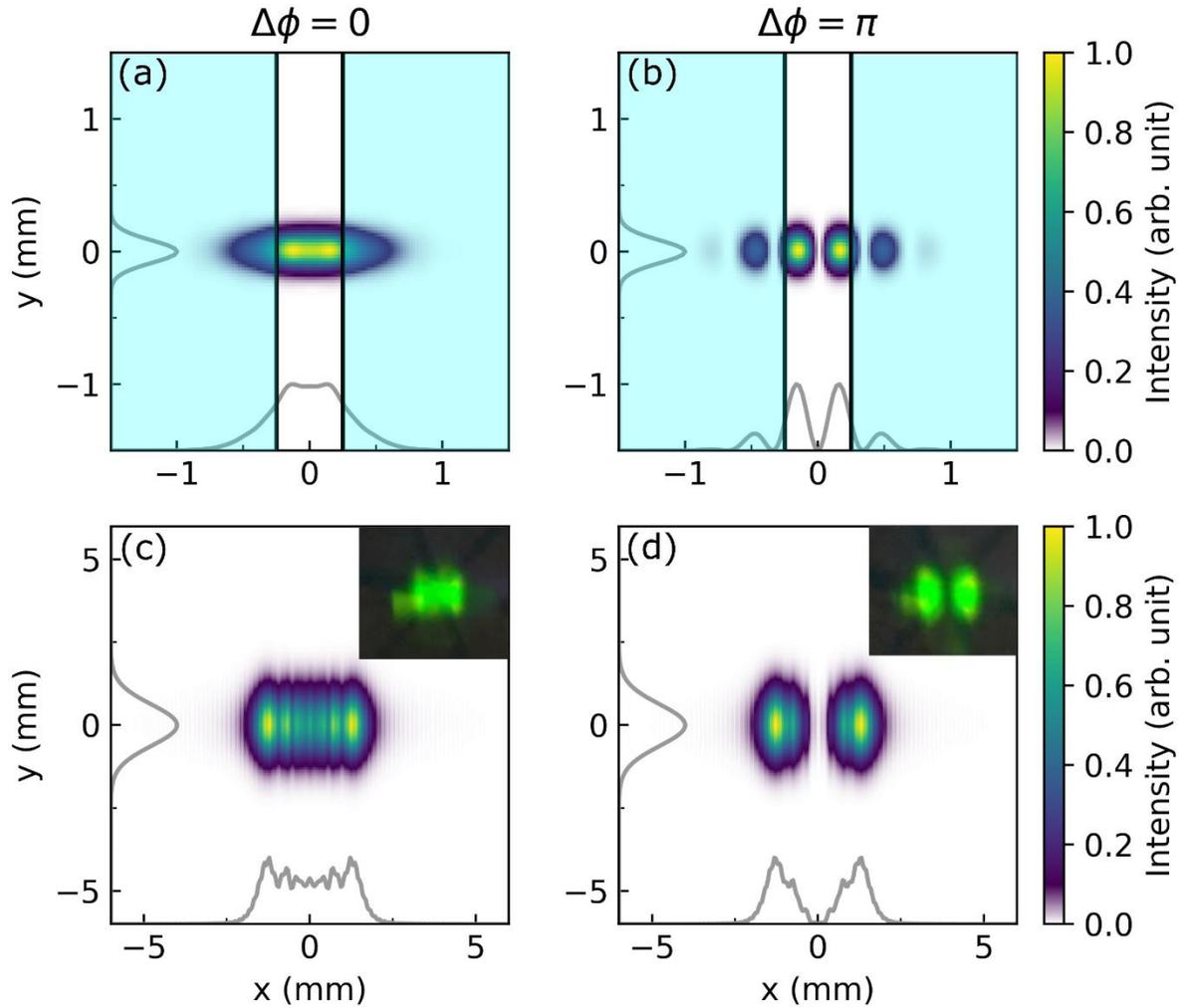

Fig 3. Fig. (a), (b) Simulated 11$^{th}$ harmonic profile on the out-coupling-mirror surface, for $\Delta\phi = 0$ and $\Delta\phi = \pi$, respectively. The shaded area is blocked by the out-coupling mirrors, and most of the harmonic power is coupled out through the gap. (c), (d) The simulated harmonic profile at a far distance away (0.7 m) from the mirror gap. Gray curves show integrated power distribution along the horizontal (x) and vertical (y) directions. Inset photos: experimentally observed 11$^{th}$ harmonic beam profiles.

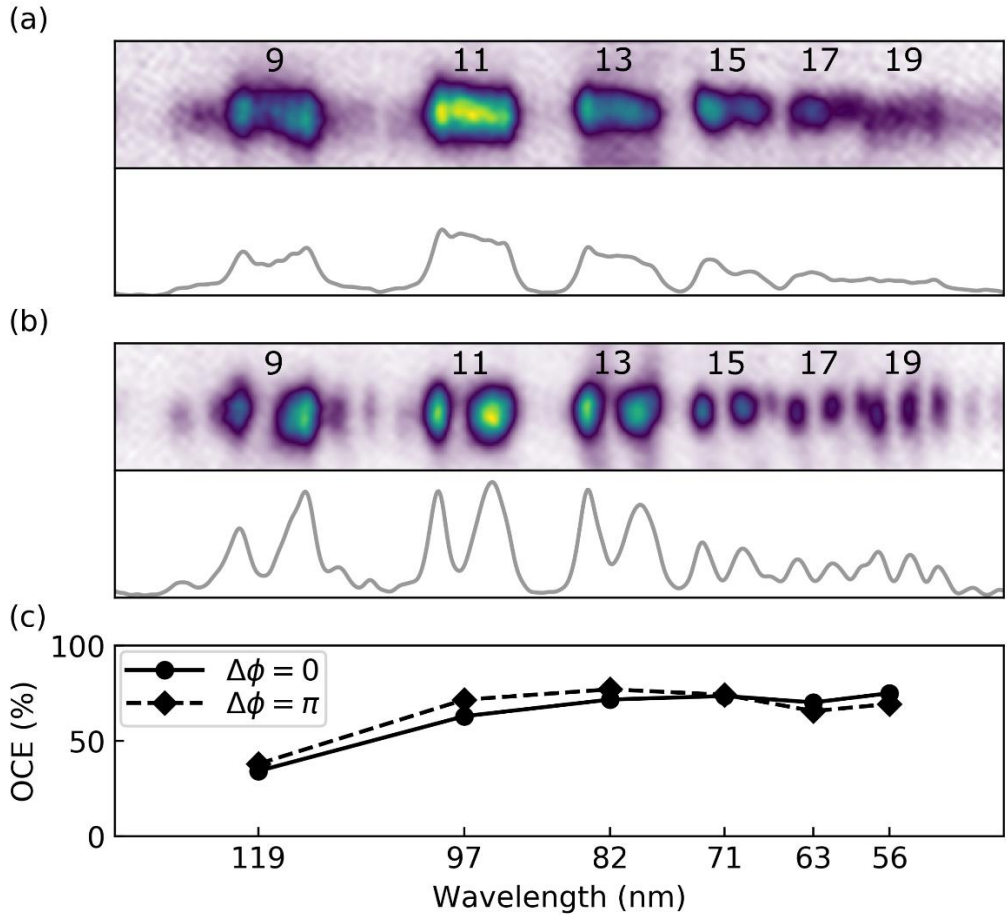

Fig 4. (a), (b) Upper panel: images of harmonics dispersed on a fluorescent plate. Lower panel: harmonic photon flux integrated in the vertical direction. Results in (a) and (b) are shown for $\Delta\phi = 0$ and $\Delta\phi = \pi$, respectively. A 2-dimensional low-pass filter in Fourier domain (not shown) is used to remove a noise pattern on the image originating from the camera. Asymmetry of the harmonics is caused by a slight misalignment between the mirror gap and the bisection axis. The images shown here are taken with the cavity locked and using pure Xe gas at room temperature. (c) Theoretically estimated out-coupling efficiencies (OCE) from the cavity for harmonic orders 9 to 19 (119 nm to 56 nm) and different $\Delta\phi$, see Supplementary [52].

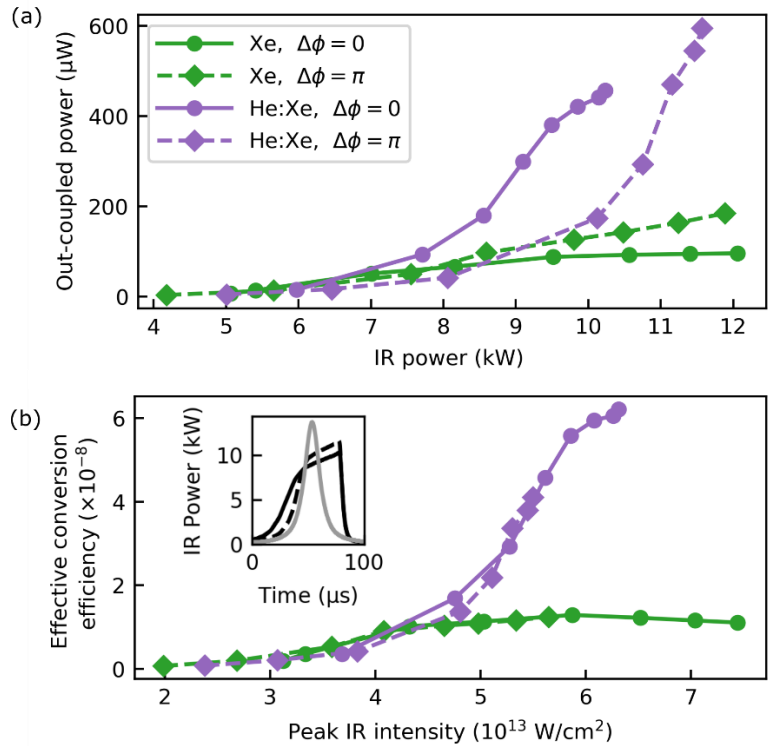

Fig 5. (a) Out-coupled 11[th] harmonic power (back-calculated to the point right after the out-coupling mirror pair, see Supplementary [52]) as a function of intracavity (single beam) fundamental drive power, shown for different generation media and relative phases. Data is taken when the cavity is swept across the resonance. As shown in Ref. [12], similar harmonic power is expected when the cavity is locked with a similar intracavity power level. (b) Effective conversion efficiency (defined in the text) as a function of peak drive intensity. Inset shows intracavity IR power when the cavity is swept across a resonance, with $\Delta\phi = 0$ (continuous black) and $\Delta\phi = \pi$ (dashed black) configurations using a He:Xe mixture gas target, displaying clear deviations from the Lorentzian lineshape obtained without a gas target (gray), indicating significant plasma density. Green traces are recorded with pure Xe with 260 kPa backing pressure at room temperature. Purple traces are recorded with 9:1 He:Xe mixture with 4100 kPa backing pressure heated to ~560 ˚C.

**Supplemental material to "Noncollinear enhancement cavity for record-high out-coupling efficiency of an extreme-UV frequency comb"**

Chuankun Zhang[1], Stephen B. Schoun[1], Christoph M. Heyl[1,*], Gil Porat[1,†], Mette B. Gaarde[2], Jun Ye[1]

[1]JILA, National Institute of Standards and Technology and Department of Physics, University of Colorado, Boulder, Colorado 80309, USA

[2]Department of Physics and Astronomy, Louisiana State University, Baton Rouge, Louisiana 70803 USA

1. **Extraction of short-trajectory HHG dipole yield and phase**

In the semi-classical model for high-harmonic generation (HHG) [1,2], electrons ionized in a strong laser field can undergo two distinct types of trajectories, known as the long trajectories and the short trajectory, before returning to the parent ion and emitting a high-energy photon. Harmonics generated from the long-trajectory contribution exhibit a fast-varying intensity-dependent phase and therefore a strong phase-front curvature at the laser focus. In the far field, the long-trajectory harmonic beams have larger divergence and create large halos. In contrast, the short-trajectory harmonic beams give on-axis emission with small beam divergence [3].

The HHG dipole yield and phase (Fig. S1(a)) used in our simulations are obtained from the solution of the time-dependent Schrödinger equation (TDSE) [4] and therefore contain both long-trajectory and short-trajectory information. In the experiment, the short-trajectory contribution is selected by choosing proper phase-matching conditions [5]. To reproduce the experimental results in our simulation, we numerically extract the short-trajectory dipole contribution from the raw TDSE results.

For well-above-threshold harmonics, one can write the intensity-dependent dipole (in atomic units) as

$$d_q(I) = \sum_j A_j e^{-i\alpha_j U_p(I)/\omega}$$

where $U_p(I) = I/4\omega^2$ is the pondermotive energy, proportional to the laser intensity I. $\omega$ is the laser frequency. Each amplitude $A_j$ and the corresponding phase coefficient $\alpha_j$ represent the contribution from a particular (quantum-mechanical) trajectory j. For the long trajectories, $\alpha_j > \pi$, which corresponds to a faster-varying phase as a function of laser intensity. As a generalization [6–8], the full quantum-mechanical HHG dipole can be written as

$$d_q(I) = \int \tilde{d}_q(\alpha) e^{-i\alpha U_p(I)/\omega} d\alpha$$

Then, for a given laser intensity $I_0$, one can perform a Fourier transform

$$\tilde{d}_q(\alpha, I_0) = \int d_q(I) e^{i\alpha U_p(I)/\omega} W(I - I_0) \, dI$$

to obtain the weight of the quantum-path distribution in a continuous phase coefficient α domain for intensities near $I_0$. $W(I - I_0)$ is a narrow window function peaked at $I_0$. For simplicity, we treat the amplitude of $d_q(I)$ as a constant in this transformation. The resulting quantum path distribution for harmonic order 17 in Xe driven with a 1070 nm laser is shown in Fig. S1(b), as an example.

In the semi-classical model [1,2], the highest (cutoff) photon energy that can be generated in an HHG process at a given laser intensity is $I_p + 3.2U_p$ where $I_p$ is the ionization potential of the atom. For harmonics with a photon energy larger than $I_p$, one can find a threshold laser intensity where the given harmonic order is exactly at the cutoff. Below this intensity, the generalized long and short trajectory distribution merge together and cannot be separated. We therefore leave the dipole data below this cutoff intensity unchanged. A numerical filter (Hann window) for intensities beyond the cutoff intensity is applied to the transformed dipole data to extract the short-trajectory contributions at $|\alpha| < \pi$.

The filtered data is transformed back to compute the intensity-dependent dipole phase. We smooth the dipole yield with a moving average to get rid of the fast oscillations arising from interferences between the different quantum paths. Resulting dipole yield and phase as well as the quantum path distribution are shown in Fig. S2.

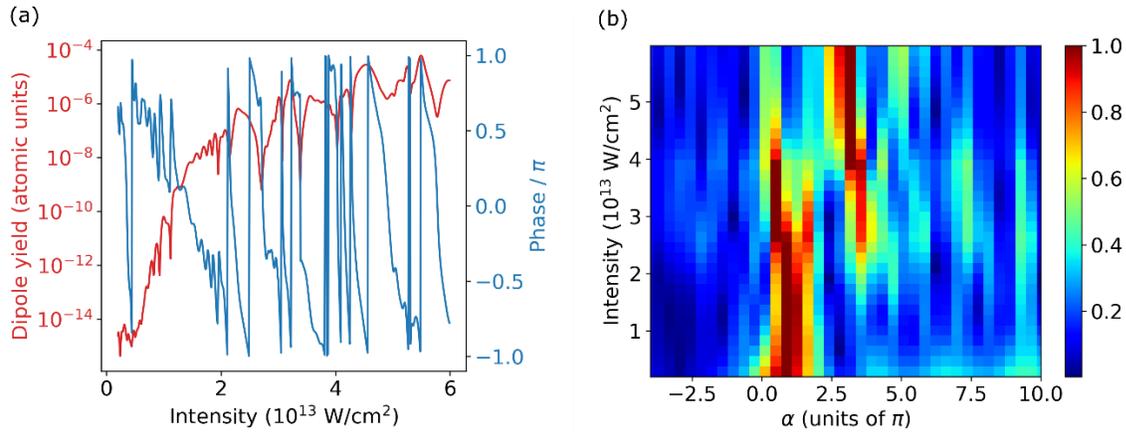

Figure S1: (a) HHG dipole $d_q(I)$ yield and phase from TDSE solutions for the 17[th] harmonic in Xe, driven by 1070 nm laser. (b) Corresponding quantum path distribution, $\tilde{d}_q(\alpha, I_0)$. We choose the 17[th] harmonic as an example because of its clear long-trajectory contribution.

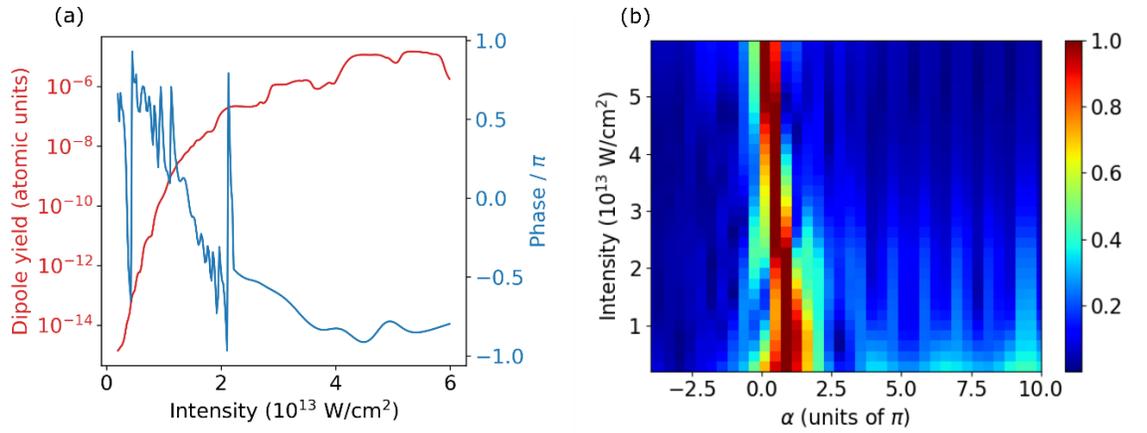

Figure S2: (a) Filtered HHG dipole yield and phase for the 17th harmonic in Xe, driven by 1070 nm laser. (b) Corresponding quantum path distribution.

**2. Harmonic power uncertainties**

The out-coupled 11th harmonic is reflected by a gold mirror and a gold grating before being measured on a NIST-calibrated detector. Here, we list the efficiency of each element in the detection system and the corresponding fractional uncertainty.

2.1 Gold Mirror

For the 11th harmonic, the angle of incidence on the gold mirror is $75 \pm 0.5°$. This gives a reflectivity of $0.61 \pm 0.01$. We take the fractional uncertainty to be $\pm 2\%$.

2.2 Gold Grating

The diffraction efficiency of the gold grating for the 11th harmonic is calculated based on Rigorous Coupled Wave Analysis. The incidence angle on the gold grating is $80.3 \pm 0.5°$, corresponding to a diffraction efficiency for the 11th harmonic of $0.307 \pm 0.007$. We take the fractional uncertainty to be $\pm 3\%$.

2.3 NIST calibrated detector

Based on the calibration data from NIST, our detector quantum efficiency is $0.07 \pm 0.01$. We take the fractional uncertainty to be $\pm 15\%$.

2.4 Total uncertainty

For the measured power of the 11th harmonic, the total fractional uncertainty comes to $\sqrt{(2\%)^2 + (3\%)^2 + (15\%)^2} = 15.5\%$.

## 3. Harmonic out-coupling efficiency (OCE)

The OCE is theoretically estimated as the ratio of the integrated XUV power through the mirror gap to the total XUV power incident on the out-coupling-mirror surface plane, see Fig. 3. The uncertainty in our estimation comes directly from experimental uncertainties in measuring the mirror gap size $d$ and the misalignment $\delta$ between the mirror gap center and the fundamental bisection axis. We measured the mirror-gap size to be $d = 0.5 \pm 0.05$ mm. We assume a possible misalignment $\delta = 0.1$ mm for the uncertainty estimation. Based on these experimental parameters, we deduce the OCE for different harmonic order $q$ and relative phase $\Delta\phi$ from our simulation, see Table S1. The same data are used in Fig. 4(c).

| $\Delta\phi$ \ $q$ | 9 | 11 | 13 | 15 | 17 | 19 |
|---|---|---|---|---|---|---|
| 0 | 34% | 63% | 72% | 74% | 70% | 75% |
| $\pi$ | 38% | 72% | 77% | 74% | 66% | 69% |

Table S1: OCE value for different harmonic order $q$ and relative phase $\Delta\phi$. We estimate an overall fractional uncertainty of about 10%.

*Present address: German Electron Synchrotron DESY, Notkestr. 85, 22607 Hamburg, Germany & Helmholtz-Institute Jena, Fröbelstieg 3, 07743 Jena, Germany

†Present address: Department of Electrical and Computer Engineering, University of Alberta, Edmonton, Alberta T6G 1H9, Canada